\documentclass[superscriptaddress,aps,preprintnumbers,amsmath,amssymb,sort&compress,nofootinbib,10pt]{revtex4-2}

\usepackage{amsfonts,amsmath,amssymb,ascmac,bm,tensor}
\usepackage{fnpct} 
\usepackage{comment}
\usepackage{ifpdf}
\usepackage{slashed}
\usepackage{color}
\usepackage[mathscr]{eucal}
\usepackage[utf8]{inputenc}
\usepackage{physics}
\usepackage{cancel}

\usepackage{mathtools}

\ifpdf
  \usepackage{graphicx}     
  \usepackage[bookmarksopen,colorlinks=true,linkcolor=bblue,citecolor=bblue,urlcolor=ppink]{hyperref}
\else     
\fi

\definecolor{ppink}{rgb}{1,0.4,0.4}
\definecolor{bblue}{rgb}{0.284602,0.317763,0.963947}

\newcommand{\GW}{\text{GW}}
\newcommand{\PBH}{\text{PBH}}

\makeatletter
\newcommand\footnoteref[1]{\protected@xdef\@thefnmark{\ref{#1}}\@footnotemark}
\makeatother

\allowdisplaybreaks[1]

\begin{document}


\preprint{KEK-TH-2535, KEK-Cosmo-0317, KEK-QUP-2023-0016, CTPU-PTC-23-28}

\title{
The Detected Stochastic Gravitational Waves \\ and Subsolar-Mass Primordial Black Holes
}

\author{Keisuke Inomata}
\affiliation{Kavli Institute for Cosmological Physics, The University of Chicago, Chicago, Illinois 60637, USA}

\author{Kazunori Kohri}
\affiliation{Division of Science, National Astronomical Observatory of Japan (NAOJ), and SOKENDAI, 2-21-1 Osawa, Mitaka, Tokyo 181-8588, Japan}
\affiliation{Theory Center, IPNS, and QUP (WPI), KEK, 1-1 Oho, Tsukuba, Ibaraki 305-0801, Japan}
\affiliation{Kavli IPMU (WPI), UTIAS, The University of Tokyo, Kashiwa, Chiba 277-8583, Japan}

\author{Takahiro Terada}
\affiliation{Particle Theory and Cosmology Group, Center for Theoretical Physics of the Universe, 
Institute for Basic Science (IBS), Daejeon, 34126, Korea}

\begin{abstract}
Multiple pulsar timing array (PTA) collaborations recently announced the evidence of common-spectral processes caused by gravitational waves (GWs). These can be the stochastic GW background and its origin may be astrophysical and/or cosmological. We interpret it as the GWs induced by the primordial curvature perturbations and discuss their implications on primordial black holes (PBHs). We show that the newly released data suggest PBHs much lighter than the Sun ($\mathcal{O}(10^{-4}) \, M_\odot$ for the delta-function curvature spectrum; $< \mathcal{O}(10^{-2})\, M_\odot$ more generally) in contrast to what was expected from the previous PTA data releases. 
\noindent
\end{abstract}

\date{\today}
\maketitle

\section{Introduction}

Recently, the evidence of the Hellings-Downs curve~\cite{Hellings:1983fr}, a smoking-gun signal of the isotropic stochastic gravitational waves (GWs) representing a particular pattern of angular correlations, has been reported by pulsar timing array (PTA) experiments, in particular, by  NANOGrav~\cite{NANOGrav:2023gor,NANOGrav:2023hde} and  by EPTA and InPTA~\cite{EPTA:2023fyk, EPTA:2023sfo, EPTA:2023xxk} (see also the results of PPTA~\cite{Reardon:2023gzh,Zic:2023gta,Reardon:2023zen} and CPTA~\cite{Xu:2023wog}).
The GWs are consistent with the stochastic GW background (SGWB), as there have not been strong hints for continuous GW signals or anisotropy~\cite{NANOGrav:2023tcn,NANOGrav:2023pdq,EPTA:2023gyr}.  A natural astrophysical interpretation of the origin of such SGWB is the superposed GW signals from binary mergers of supermassive black holes as discussed in the above PTA papers.  

Alternatively, the observed GWs may have a cosmological origin. This could be a great observational window to study the early-Universe cosmology. 
The common spectrum process was already observed in the NANOGrav 12.5-year data~\cite{NANOGrav:2020bcs} and IPTA data release 2~\cite{Antoniadis:2022pcn}, though the evidence of the Hellings-Downs curve was not observed at that time. 
Since then, the cosmological GW sources for the PTA experiments have been enthusiastically studied, e.g., in the context of topological defects such as cosmic strings and domain walls~\cite{Ellis:2020ena,Datta:2020bht,Samanta:2020cdk,Buchmuller:2020lbh, Blanco-Pillado:2021ygr,Ferreira:2022zzo, Bian:2022qbh}, cosmological first-order phase transitions~\cite{NANOGrav:2021flc,Xue:2021gyq,Nakai:2020oit,DiBari:2021dri,Sakharov:2021dim,Li:2021qer,Ashoorioon:2022raz,Benetti:2021uea,Barir:2022kzo,Hindmarsh:2022awe}, scalar-induced GWs associated with primordial black holes (PBHs)~\cite{Vaskonen:2020lbd,DeLuca:2020agl,Kohri:2020qqd,Domenech:2020ers,Papanikolaou:2020qtd,Inomata:2020xad,Kawasaki:2021ycf,Dandoy:2023jot}, and inflationary GWs~\cite{Vagnozzi:2020gtf, Li:2020cjj, Bhattacharya:2020lhc, Kuroyanagi:2020sfw} (see also Ref.~\cite{Madge:2023dxc} for a comprehensive study on the cosmological sources).

After the recent announcements, a variety of explanations of the SGWB were proposed: 
cosmic strings~\cite{Ellis:2023tsl, Wang:2023len, King:2023cgv}, 
domain walls~\cite{Guo:2023hyp, Kitajima:2023cek, Bai:2023cqj, King:2023cgv}, 
a first-order phase transition~\cite{Zu:2023olm, Han:2023olf, Megias:2023kiy, Fujikura:2023lkn}, 
inflation (first-order GWs)~\cite{Vagnozzi:2023lwo}, 
second-order (scalar-induced) GWs and PBHs~\cite{Franciolini:2023pbf, Cai:2023dls}, 
parametric resonance in the early dark energy model~\cite{Kitajima:2023mxn}, 
turbulence due to the primordial magnetic field~\cite{Li:2023yaj}, 
axion-like particles and gravitational atoms~\cite{Yang:2023aak}. In addition, the importance of softening of the equation of state by the QCD crossover was pointed out~\cite{Franciolini:2023wjm}, the effects of the SGWB on neutrino oscillations were discussed~\cite{Lambiase:2023pxd}, and 
probing dark matter density was discussed in the context of supermassive binary BH mergers~\cite{Ghoshal:2023fhh}.

In this work, we discuss the implications of the recent observation of the stochastic GWs by the PTA experiments on PBHs. 
PBHs have been attracting a lot of attention because of their potential to explain dark matter and/or the binary BH merger signals detected by LIGO-Virgo-KAGRA  Collaborations~\cite{Bird:2016dcv,Clesse:2016vqa,Sasaki:2016jop,Kashlinsky:2016sdv,Kashlinsky:2018mnu, Garcia-Bellido:2020pwq,Garcia-Bellido:2020pwq} (see Refs.~\cite{Sasaki:2018dmp,Carr:2020gox,Green:2020jor,Escriva:2022duf} for reviews). 
PBHs can be produced when large density perturbations enter the Hubble horizon. 
In general, large density perturbations can produce not only PBHs but also GWs through the nonlinear interaction~\cite{Ananda:2006af,Baumann:2007zm}. 
These scalar-induced GWs are an important probe of PBHs~\cite{Saito:2008jc,Saito:2009jt,Inomata:2016rbd,Ando:2017veq,Espinosa:2018eve,Kohri:2018awv,Cai:2018dig,Bartolo:2018evs,Bartolo:2018rku,Unal:2018yaa,Byrnes:2018txb,Inomata:2018epa,Clesse:2018ogk,Cai:2019amo,Cai:2019jah, Chen:2019xse} (see also Refs.~\cite{Domenech:2021ztg, Yuan:2021qgz} and references therein). 
In particular, after the first detection of GWs from the merger of $\sim 30 \, M_\odot$ BHs, the connection between the PTA experiments and the scalar-induced GW signals associated with $\mathcal O(10) \, M_\odot$ PBHs have been focused on in Refs.~\cite{Inomata:2016rbd,Orlofsky:2016vbd,Nakama:2016gzw,Garcia-Bellido:2017aan,Di:2017ndc,Ando:2017veq,Cheng:2018yyr,Kohri:2018qtx,Chen:2019xse}.

\section{The Gravitational-Wave Signals and induced gravitational waves}

The SGWB detected by the PTA collaborations may be explained by cosmological (New Physics) GWs. The New Physics interpretations of the SGWB were studied by the NANOGrav Collaboration~\cite{NANOGrav:2023hvm} and by the EPTA/InPTA Collaboration~\cite{EPTA:2023xxk}. In particular, they studied the interpretation of the data by the induced GWs.    
We first review their results.

The cosmological abundance of the GWs is conventionally parametrized by $\Omega_\text{GW}(f) = \rho_\text{GW}(f)/\rho_\text{total}$, where $\rho_\text{GW}(f)$ is related to the total energy density of the GWs by $\rho_\text{GW} = \int \mathrm{d}\ln f \rho_\text{GW}(f)$.  The GW spectrum in the PTA experiments is parametrized as~\cite{Arzoumanian:2018saf} 
\begin{align}
    \Omega_\text{GW} = \frac{2\pi^2 f_*^2}{3 H_0^2} A_\text{GWB}^2 \left( \frac{f}{f_*} \right)^{5- \gamma}, 
\end{align}
where $H_0$ is the Hubble parameter, $A_\text{GWB}$ the amplitude of the GWs, $f_*$ the pivot-scale frequency often adopted as $f_* = f_\text{yr} = (1 \, \mathrm{yr})^{-1}\approx 32 \, \mathrm{nHz}$, and $\gamma$ the power index of the spectrum. 

Fig.~1 (b) in Ref.~\cite{NANOGrav:2023gor} shows that 
the data favor $\gamma \approx 3$ ($\gamma = 3.2 \pm 0.6$ (90\% credible region)~\cite{NANOGrav:2023gor}), compared to $\gamma \approx 2$ or $\approx 4$, corresponding to the power-law index $5-\gamma \approx 2$. For simplicity, we assume an integer value $\gamma = 3$.

There are several ways to interpret this power law in terms of the induced GWs. 
\begin{itemize}
\item  The power-law GWs can be explained by the power-law curvature perturbations $\mathcal{P}_\zeta (k) \propto k$ in the relevant range of frequencies. With such $\mathcal{P}_\zeta(k)$, the induced GWs approximately behave as $\Omega_\text{GW} \sim \mathcal{P}_\zeta^2 \propto f^2$.  This requires a nontrivial condition on the underlying inflation model so that $\mathcal{P}_\zeta \propto k$.  Since this is not a generic consequence of inflation models, we do not focus on this case in this paper. A recent discussion based on inflation models can be found in Ref.~\cite{Franciolini:2023pbf}.

\item The spectral slope can be interpreted as the so-called universal infrared (IR) tail when the curvature perturbations are sufficiently steep. For the induced GWs produced during the radiation-dominated (RD) era, the universal IR tail has the $\Omega_\text{GW} \propto f^3$ slope~\cite{Cai:2019cdl} for generic underlying curvature perturbations up to logarithmic corrections~\cite{Yuan:2019wwo}.  However, this slope becomes $f^2$ under some conditions.
    \begin{itemize}
        \item When the induced GWs are produced in a cosmological era with the equation-of-state parameter $w$, the power-law index of the universal IR tail was worked out to be $\Omega_\text{GW} \propto f^{3 - 2|\frac{1-3w}{1+3w}|}$~\cite{Domenech:2020kqm}.  In particular, the power is 2 as desired if $w = 1$ or $1/9$.  For example, the $w=1$ case is realized when the Universe is dominated by the kinetic energy of a scalar field. In the remainder of this paper, we focus on the standard RD era, so we do not consider this option. 
        
        \item Even in the RD era, the $\Omega_\text{GW} \sim f^2$ scaling can be realized for an extended range of the frequencies when the spectrum is narrow~\cite{Pi:2020otn}.  In particular, the limit of the delta function $\mathcal{P}_\zeta$ leads to the $\Omega_\text{GW} \sim f^2$ scaling without the restriction on the frequency range. The delta function peak is, of course, not physical nor realistic, but it would approximate sufficiently peaked spectra.\footnote{
        More precisely, the ratio of the peak frequency $f_\text{p}$ and the spectral break frequency $f_\text{b}$ that divides the $f^3$ and $f^2$ scalings is given by the width of the power spectrum of the curvature perturbations up to an $\mathcal{O}(1)$ factor when the peak is narrow~\cite{Pi:2020otn}.  When the peak is broad, the GW spectrum also becomes broad and the IR tail is given by the universal $f^3$ slope. See also Appendix~\ref{sec:finite_width}. \label{fn:IR_scaling_change}
        }
        
    \end{itemize}
\end{itemize}
The last item above, i.e., the delta function power spectrum was studied in Refs.~\cite{Saito:2008jc,NANOGrav:2023hvm, EPTA:2023xxk} along with other example spectra such as the log-normal function and the top-hat box-shaped function~\cite{Saito:2009jt}.  Because of the simplicity and the guaranteed $f^2$ scaling, we discuss the delta function power spectrum $\mathcal{P}_\zeta$ as a first step to interpret the PTA GW signals in terms of the induced GWs. We briefly discuss in Appendix~\ref{sec:finite_width} how the result changes when a finite width of the peak of the power spectrum is taken into account. 

Let us here summarize the equations for the induced GWs. In the Newtonian gauge, the metric perturbations are given by 
\begin{align}
    \dd^2 s =& -a^2\left[ (1+2\Phi) \dd \eta^2 + \left((1 - 2 \Psi)\delta_{ij} +\frac{h_{ij}}{2} \right) \dd x^i \dd x^j\right],
    \label{eq:newton_gauge}
\end{align}
where $\Phi$ and $\Psi$ are the scalar perturbations and $h_{ij}$ is the tensor perturbation, which describes GWs.
We have neglected vector perturbations, because we focus on the GWs induced by the scalar perturbations throughout this work.
In the following, we consider the perfect fluid, which enables us to take $\Psi = \Phi$. 
Then, from the Einstein equation, the equation of motion of the tensor perturbations is given by 
\begin{align}
{h^{\lambda}_{\bm{k}}}''\!(\eta) + 2 \mathcal H {h^{\lambda}_{\bm{k}}}'\!(\eta) + k^2 h^{\lambda}_{\bm{k}}(\eta) = 4 S^{\lambda}_{\bm{k}}(\eta),
\label{eq:h_eom}
\end{align}
where $\bm{k}$ ($k = |\bm{k}|$) and $\lambda$ denote the wavenumber and the polarization of the tensor perturbations, the prime denotes $\partial/\partial \eta$, and $\mathcal H = a'/a$ is the conformal Hubble parameter. 
The source term $S^\lambda_{\bm k}$ is given by
\begin{align}
\label{eq:s}
S^\lambda_{\bm{k}} =&\,  \int \! \frac{\dd^3 q}{(2\pi)^3} \, e^{\lambda}_{ij}(\hat{\bm k})q^i q^j \bigg[ 2 \Phi_{\bm{q}} \Phi_{\bm{k-q}}  
 + \frac{4}{3(1+w)}   \left(\mathcal{H}^{-1}\Phi'_{\bm{q}} + \Phi_{\bm{q}} \right)
 \left( \mathcal{H}^{-1}\Phi'_{\bm{k-q}} + \Phi_{\bm{k-q}}\right) \bigg],
\end{align}
where $e^\lambda_{ij}(\hat{\bm k})$ is the polarization tensor and $w = 1/3$ in an RD era. 
By solving the equation of motion and taking the late-time limit during an RD era, we obtain the energy density parameter of the induced GWs during an RD era as 
\begin{align}
\label{eq:omega_gw}
\tilde \Omega_{\GW}\left(f\right) = \int_0^\infty \textrm{d}v \int_{\left|1-v\right|}^{1+v} \textrm{d}u\: \mathcal{K}\left(u,v\right) \mathcal{P}_\mathcal{\zeta}\left(uk\right) \mathcal{P}_\mathcal{\zeta}\left(vk\right) \,,
\end{align} 
where $f=k/(2\pi)$ is the frequency of the GW and $\mathcal{P}_{\mathcal{\zeta}}$ is the power spectrum of the curvature perturbation, and the integration kernel $\mathcal{K}$ is given by~\citep{Espinosa:2018eve,Kohri:2018awv}
\begin{align}
\mathcal{K}\left(u,v\right) = \frac{3\left(4v^2-(1+v^2-u^2)^2\right)^2\left(u^2+v^2-3\right)^4}{1024\,u^8 v^8}\left[\left(\ln\left|\frac{3-(u+v)^2}{3-(u-v)^2}\right| - \frac{4uv}{u^2+v^2-3}\right)^2 + \pi^2 \Theta(u+v-\sqrt{3}) \right] \,.
\end{align}
Note that the energy density parameter of the induced GWs during an RD era asymptotes to $\tilde \Omega_\GW$ after the peak-scale perturbations enter the horizon.

Taking into account the following matter-dominated and dark-energy-dominated eras, the current energy density of the induced GWs is given by 
\begin{align}
    \Omega_\GW h^2 = 
    0.43 \left( \frac{g_*}{80}\right) \left( \frac{g_{*,s}}{80}\right)^{-4/3} \Omega_r h^2 \tilde \Omega_\GW,
\end{align}
where $g_*$ and $g_{*, s}$ are the effective relativistic degrees of freedom for the energy and entropy densities, respectively, when the GW energy density parameter becomes constant during the RD era, and $\Omega_r$ is the current energy density parameter of radiation ($\Omega_r h^2 \simeq 4.2 \times 10^{-5}$). See Ref.~\cite{Saikawa:2018rcs} for the temperature dependence of $g_*(T)$ and $g_{*, s}(T)$. 

As motivated above, we consider the monochromatic (delta function) power spectrum given by 
\begin{align}
    \mathcal P_\zeta = A_\zeta\, \delta(\ln(k/k_*)), \label{eq:P_zeta_delta}
\end{align}
where $A_\zeta$ governs the overall normalization and $k_*$ is the wavenumber at which there is a spike in the power spectrum. 
In this case, the energy density parameter of the induced GWs [Eq.~(\ref{eq:omega_gw})] can be expressed as 
\begin{align}
\tilde \Omega_{\text{GW}}(f)=& \frac{3 A_\zeta^2}{64}  \left(\frac{4-\tilde f^2}{4} \right)^2  \tilde f^2 \left(3 \tilde f^2-2\right)^2 \nonumber \\
&  \times  \left( \pi^2 (3 \tilde f^2-2)^2 \Theta (2\sqrt{3}-3 \tilde f) + \left( 4+(3\tilde f^2-2) \log \left| 1- \frac{4}{3 \tilde f^2} \right| \right)^2 \right) \Theta (2-\tilde f), \label{Omega_GW_RD_delta}
\end{align}
where the dimensionless wavenumber $\tilde{f}\equiv f/f_*$ is introduced for notational simplicity.  In the IR limit $\tilde{f} \ll 1$, the spectrum reduces to 
\begin{align}
    \tilde \Omega_\text{GW} \simeq \frac{3 A_\zeta^2}{4} \tilde{f}^2 \left(\pi^2 + \left( 2 + \log\frac{3\tilde{f}}{4}  \right)^2 \right).
\end{align}
Indeed, this scales as $f^2$ up to the logarithmic correction that becomes more and more important in the limit $\tilde{f} \ll 1$. 

\begin{figure}[tbh]
    \centering
    \includegraphics[width=0.6\textwidth]{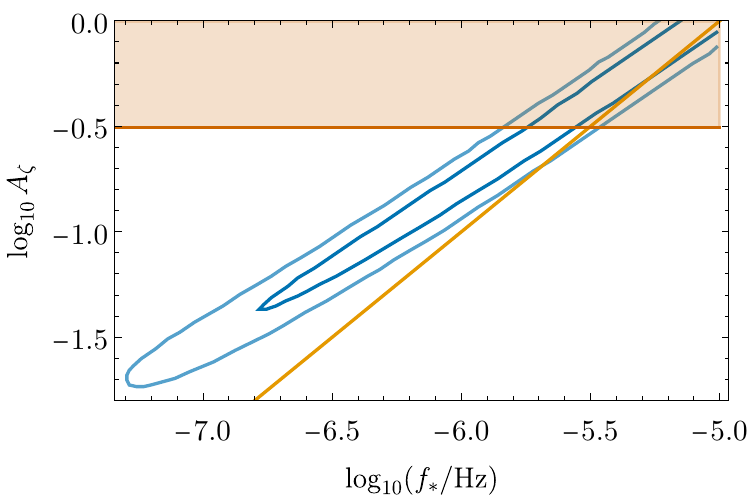}
    \caption{The favored parameter space (blue contours) for the delta function power spectrum taken from Ref.~\cite{NANOGrav:2023hvm}.
    The darker (lighter) blue contour corresponds to the $68\%$ ($95\%$) Bayesian credible region. 
    The orange line illustrates the approximate parameter degeneracy relation $A_\zeta \propto f_*$ [Eq.~\eqref{eq:degeneracy}]. The vermilion-shaded region is the dark radiation constraint from the big-bang nucleosynthesis, $\Omega_\text{GW}h^2 < 1.8 \times 10^{-6}$~\cite{Kohri:2018awv}. 
    }
    \label{fig:f-A}
\end{figure}

Note that the above spectrum depends on the combination $A_\zeta / f_*$ up to the logarithmic correction, so one must expect the parameter degeneracy in the direction $A_\zeta \propto f_*$.  This is consistent with the analyses by the PTA collaborations~\cite{NANOGrav:2023hvm, EPTA:2023xxk}. The NANOGrav result~\cite{NANOGrav:2023hvm} is shown by blue contours in Fig.~\ref{fig:f-A}. In this figure, the orange line for comparison shows a linear relation
\begin{align}
    A_\zeta = 10^{-2} \left(\frac{f_*}{10^{-7} \, \mathrm{Hz}}\right). \label{eq:degeneracy}
\end{align}
As the characteristic frequency approaches the NANOGrav frequency range $[2 \times 10^{-9} \, \mathrm{Hz}, \, 6 \times \, 10^{-8}\, \mathrm{Hz}]$, the relevant part of the GW spectrum ceases to be the IR tail, which has the $f^2$ scaling.  This is why the deviation of the orange straight line from the blue contours becomes larger toward the left part of the figure. The vermilion-shaded region is excluded by the dark radiation constraint $\Omega_\mathrm{GW}h^2 < 1.8 \times 10^{-6}$ from the big-bang nucleosynthesis~\cite{Kohri:2018awv} because of the overproduction of GWs. See also Ref.~\cite{Clarke:2020bil} for a slightly stronger constraint $\Omega_\text{GW}h^2 < 1.7 \times 10^{-6}$ from  the cosmic microwave background.

Already at this stage, the degeneracy allows the parameter space to extend to much higher frequencies than the nanohertz ballpark. This indicates that the new data analyses by the PTA collaborations prefer smaller-mass PBHs than expected so far.  

\section{Implications for primordial black holes}

In this section, we show that subsolar-mass PBHs are favored by the new data.  To this end, we summarize the formulas for PBHs.  
When it comes to the abundance of PBHs, there is a huge uncertainty depending on the calculation scheme. 
For definiteness, we basically adopt the method used by the NANOGrav analyses~\cite{NANOGrav:2023hvm}.  In particular, our prescription here (as in Ref.~\cite{NANOGrav:2023hvm}) is based on Carr's formula (also known as the Press-Schechter formalism)~\cite{Carr:1975qj}, but the result changes significantly when one adopts the peaks theory, within which there are varieties of methods with varying results~\cite{1986ApJ...304...15B, Yoo:2018kvb, Germani:2018jgr, Young:2019osy, Suyama:2019npc, Germani:2019zez, Young:2020xmk, Yoo:2020dkz,Musco:2020jjb,Kitajima:2021fpq,Young:2022phe,DeLuca:2023tun}.  There are also uncertainties on the choice of the window function and the critical density (shortly introduced below), on which we will come back to in Sec.~\ref{sec:conclusion}.

PBHs are formed shortly after an extremely enhanced curvature perturbation enters the Hubble horizon~\cite{Hawking:1971ei, Carr:1974nx, Carr:1975qj}. 
Therefore, the mass of PBHs is related to the wavenumber of the perturbations that produce PBHs:
 \begin{align}
     M = \gamma \, M_H 
     \simeq & \, 6.1 \times 10^{-4} \, M_{\odot} \left(\frac{\gamma}{0.2}\right) \left(\dfrac{g_*(T)}{80}\right)^{1/2}\left(\dfrac{g_{*,s}(T)}{80}\right)^{-2/3}\left(\dfrac{6.5\times 10^{7} \,\text{Mpc}^{-1}}{k}\right)^{2} \nonumber \\
     \simeq & \, 6.1 \times 10^{-4} \, M_{\odot} \left(\frac{\gamma}{0.2}\right)\left(\dfrac{g_*(T)}{80}\right)^{1/2}\left(\dfrac{g_{*,s}(T)}{80}\right)^{-2/3}\left(\dfrac{1.0\times 10^{-7} \,\text{Hz}}{f}\right)^{2},
 \end{align}
where $\gamma$ is the ratio between the PBH mass and the horizon mass, for which we take $\gamma = 0.2$ as a fiducial value~\cite{Carr:1975qj}, and $T$ is the temperature at the PBH production.

The PBH abundance per log bin in $M$ is given by \citep{Ando:2018qdb}
\footnote{The tilde on $\tilde{f}_\text{PBH} (M)$ is introduced to distinguish it with its integrated quantity $f_\text{PBH}$. 
}
\begin{align}
\tilde f_\PBH\left(M\right)\simeq 
\gamma^{3/2}\left(\dfrac{\beta(M)}{2.6\times10^{-9}}\right)\left(\dfrac{80}{g_*(T)}\right)^{1/4}\left(\dfrac{0.12}{\Omega_{\text{DM}}h^2}\right)\left(\dfrac{M_{\odot}}{M}\right)^{1/2}.
\end{align} 
The production rate $\beta$ is given by~\cite{Carr:1975qj} 
\begin{align}     \beta(M)=\int_{\delta_c}\dfrac{d\delta}{\sqrt{2\pi}\,\sigma(M)}\exp\left(-\dfrac{\delta^2}{2\,\sigma^2(M)}\right)\simeq \dfrac{\sigma(M)}{{\sqrt{2\pi}}\,\,\delta_c}\exp\left(-\dfrac{\delta_c^2}{2\,\sigma^2(M)}\right),
\end{align}
where $\delta_c$ is the threshold value of the overdensity for PBH production.
As a fiducial value, we take $\delta_c = 0.45$~\cite{Harada:2013epa, Musco:2004ak,Polnarev:2006aa,Musco:2012au,NANOGrav:2023hvm}. 
The $\sigma$ is the coarse-grained density contrast, given by 
 \begin{align}
      \sigma^2(k)= \dfrac{16}{81}\int \frac{dq}{q}\:\left(\dfrac{q}{k}\right)^4 \, W^2\left(\dfrac{q}{k}\right) \, \mathcal T^2\left(q,k^{-1}\right) \, \mathcal{P}_{\zeta}(q) \,. \label{eq:sigma2}
 \end{align}
 Here, $W(x)$ is a window function, which we take $W(x)=e^{-x^2/2}$, and $\mathcal T$ is the transfer function of the density perturbations during an RD era:
 \begin{align}
\mathcal T\left(q,k^{-1}\right)= 3 \left[\sin\left(\frac{q}{\sqrt{3}k}\right)-\left(\frac{q}{\sqrt{3}k}\right) \cos\left(\frac{q}{\sqrt{3}k}\right)\right]\big/\left(\frac{q}{\sqrt{3}k}\right)^3 \,.
 \end{align}
The total abundance of the PBHs can be expressed as 
\begin{align}
f_\PBH =  \int \frac{dM}{M}\, \tilde f_\PBH(M) \,.
\end{align}

\begin{figure}[tbh]
    \centering
    \includegraphics[width=0.6\textwidth]{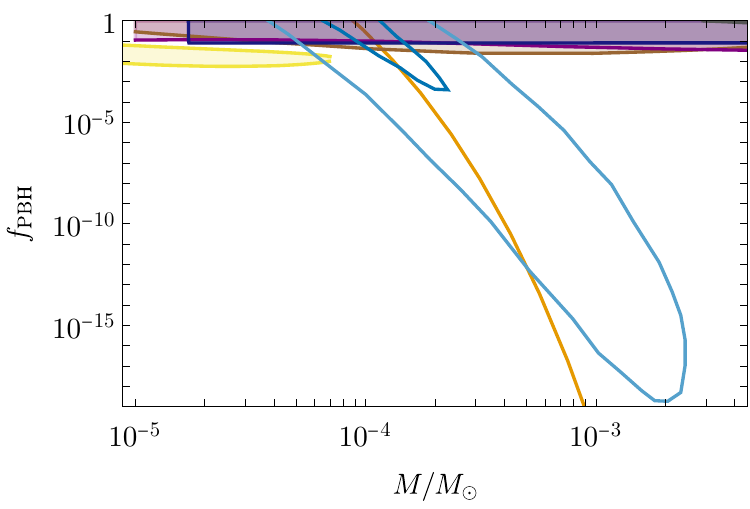}
    \caption{
    The favored region (blue contours) on the $M$-$f_\text{PBH}$ plane. 
    The dark and light blue contours correspond to the $68\%$ and $95\%$ credible regions, respectively. 
    This shows that substantially lighter PBHs than solar mass are favored. The orange curve is an approximate formula [Eq.~\eqref{eq:f(M)_approximate_formula}] with $\nu = 0.2$. The yellow shaded region near the top left corner is the 95\% allowed region~\cite{Niikura:2019kqi} to explain the ultrashort-timescale microlensing events observed by OGLE \cite{2017Natur.548..183M} combined with the Hyper Suprime-Cam (HSC) constraint~\cite{Niikura:2017zjd}.  The dark shaded region on top of the figures represents constraints on PBHs in this mass range: EROS-2~\cite{EROS-2:2006ryy} and MACHO~\cite{Macho:2000nvd} (purple), OGLE~\cite{2017Natur.548..183M, Niikura:2019kqi} (brown), caustic crossing~\cite{Oguri:2017ock} (blue), and Advanced LIGO~\cite{Wang:2016ana, Chen:2019irf, LIGOScientific:2019vic} (gray; top right corner).
 }
    \label{fig:M-fPBH}
\end{figure}

In the case of the monochromatic power spectrum, $\sigma^2(k)$ can be integrated analytically:
\begin{align}
    \sigma^2(k) = 16 A_\zeta e^{-1/\tilde{k}^2} \left( \cos^2 \left(\frac{1}{\sqrt{3}\tilde{k}} \right) + \tilde{k} \left( 3 \tilde{k} \sin^2 \left( \frac{1}{\sqrt{3}\tilde{k}} \right) - \sqrt{3} \sin \left( \frac{2}{\sqrt{3}\tilde{k}}\right) \right) \right),
\end{align}
where $\tilde{k} \equiv k / k_*$. Using these equations, we can map the contours in Fig.~\ref{fig:f-A} onto a PBH parameter space $(M, f_\text{PBH})$ numerically. Before doing it, let us obtain an approximate analytic formula to have some intuition.  
Combining the above equations and the parameter degeneracy relation Eq.~\eqref{eq:degeneracy}, we can map the degeneracy relation onto the $M$-$f_\text{PBH}$ plane: 
\begin{align}
    f_\text{PBH} \sim \tilde f_\text{PBH}(M) \sim 5.6 \times 10^{10} \frac{\nu^{1/2} }{ \delta_c} \left( \frac{M}{6.1 \times 10^{-8}M_\odot}\right)^{-3/4}  \exp \left( - \frac{\delta_c^2}{2 \nu } \left( \frac{M}{6.1 \times 10^{-8}M_\odot} \right)^{1/2} \right), \label{eq:f(M)_approximate_formula}
\end{align}
where we have approximated $\sigma^2$ as $\sigma^2 (k(M)) \sim \nu A_{\zeta}$.  The coefficient $\nu$ is introduced to show the sensitivity of $f_\text{PBH}$ on the overall normalization of $\sigma^2$.  In Fig.~\ref{fig:M-fPBH}, we set $\nu = 0.2$. The slope does not fit perfectly because the GW spectrum does not have a perfect $f^2$ scaling. This equation is not a rigorous fit but just a rough guide to the location of the contours mapped from Fig.~\ref{fig:f-A} into Fig.~\ref{fig:M-fPBH}. 

The blue contours in Fig.~\ref{fig:M-fPBH} are the main result of this paper. This shows that the PBHs much lighter than the Sun are favored by the new PTA data analysis results. If the PBH abundance is significant, they have masses of the order of $10^{-4} \, M_\odot$. More generally, the contours show the range [$5 \times 10^{-5} \leq M/M_\odot \leq 2 \times 10^{-3}$].  It should be emphasized that this is based on the assumption of the delta function $\mathcal{P}_\zeta$, but our conclusion does not qualitatively change even if we take into account the finite width of the realistic power spectrum.  This statement is backed up in Appendix~\ref{sec:finite_width}, where we study the effects of a finite width of the power spectrum of curvature perturbations (see also Ref.~\cite{NANOGrav:2023hvm}).  We find that the PBH mass can be as large as $2 \times 10^{-2}\, M_\odot$, but it is around $\mathcal{O}(10^{-4})$ and $\mathcal{O}(10^{-3})$ solar mass when the PBH abundance is large enough to be cosmologically relevant. 

The shaded regions in the upper part in Fig.~\ref{fig:M-fPBH} show the existing observational constraints on the abundance of PBHs. We see that there is a lower as well as an upper bound on the mass of the PBHs. In other words, the extension to the degeneracy direction is limited by the overproduction of PBHs.  Again, the quantitative values of these bounds  change when we drop off the assumption of the delta function curvature spectrum, but our conclusion is intact at least when the power spectrum has a narrow peak. See the analyses of NANOGrav~\cite{NANOGrav:2023hvm} and EPTA/InPTA~\cite{EPTA:2023xxk} and Appendix~\ref{sec:finite_width} for the curvature spectra with a finite width.

\section{Conclusion and Discussion}\label{sec:conclusion}

\begin{figure}[tbh]
    \centering
    \includegraphics[width=0.6\textwidth]{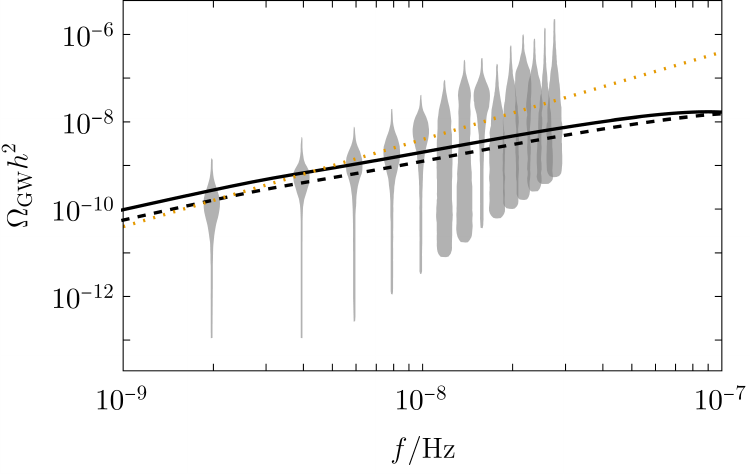}
    \caption{Comparison of the induced GW spectrum and the 14 lowest-frequency bins~\cite{NANOGrav:2023gor} of NANOGrav 15-year data~\cite{NANOGrav:2023hvm}.  
    The solid black curve represents the induced GW spectrum for the case with $M = 1.2 \times 10^{-4} \, M_\odot$ and $f_\text{PBH} = 2 \times 10^{-2}$ (corresponding to the solid black lines in Fig.~\ref{fig:prospect} below),
    whereas the dashed black curve can explain the OGLE$+$HSC data~\cite{2017Natur.548..183M, Niikura:2017zjd, Niikura:2019kqi}: $M = 6.5 \times 10^{-5} \, M_\odot$ and $f_\text{PBH} = 1.5 \times 10^{-2}$. The pure $f^2$ scaling is shown by the orange dotted line with $A_\text{GWB}|_{f_*=f_\text{yr}} = 10^{-14.1}$ . }
    \label{fig:Omega_GW}
\end{figure}

\begin{figure}[tbhp!]
    \centering
    \includegraphics[width=0.8\textwidth]{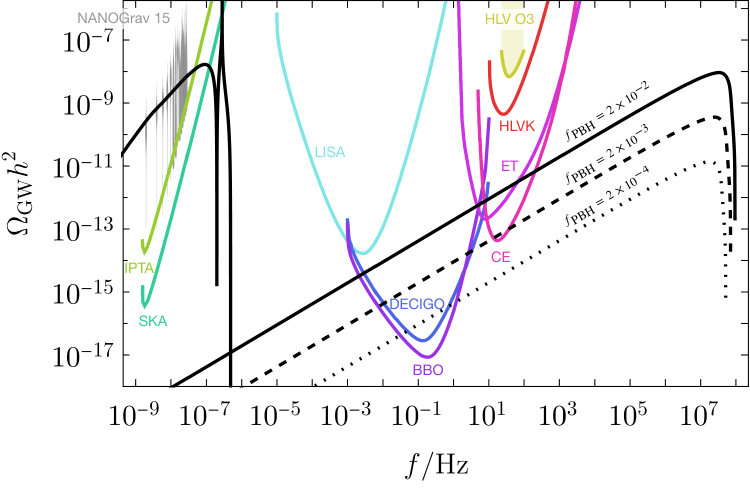} 
    \caption{ Future prospect of testing the scenario by detecting the merger-based SGWB (three black lines on the right side).  
    The spectrum depends on $M$ and $f_\text{PBH}$, but it mostly depends on the latter. From top to bottom, $f_\text{PBH} = 2 \times 10^{-2}$ (solid line), $2 \times 10^{-3}$ (dashed line), and $2 \times 10^{-4}$ (dotted line). The PBH mass was chosen within the blue contours in Fig.~\ref{fig:M-fPBH}. From top to bottom, $M/M_\odot = 1.2 \times 10^{-4}$ (solid line), $1.6 \times 10^{-4}$ (dashed line), and $2.2 \times 10^{-4}$ (dotted line). 
    The associated induced GW signal is also shown around and above the nanohertz frequency band by the black solid line. 
    The gray violin plot shows the 14 lowest-frequency bins~\cite{NANOGrav:2023gor} of the NANOGrav 15-year data~\cite{NANOGrav:2023hvm}. 
The shaded yellow region is the LIGO/Virgo O3 constraint on SGWB~\cite{KAGRA:2021kbb}. 
The colored lines are power-law-integrated sensitivity curves for the SGWB search of IPTA~\cite{Hobbs:2009yy, Manchester:2013ndt}, SKA~\cite{Carilli:2004nx, Janssen:2014dka, Weltman:2018zrl}, LISA~\cite{LISA:2017pwj, Baker:2019nia}, DECIGO~\cite{Seto:2001qf, Yagi:2011wg, Isoyama:2018rjb, Kawamura:2020pcg}, BBO~\cite{Crowder:2005nr, Corbin:2005ny, Harry:2006fi}, ET~\cite{Punturo:2010zz, Hild:2010id, Sathyaprakash:2012jk, Maggiore:2019uih}, CE~\cite{LIGOScientific:2016wof, Reitze:2019iox}, and a combination of advanced LIGO~\cite{Harry:2010zz, LIGOScientific:2014pky}, advanced Virgo~\cite{VIRGO:2014yos}, and KAGRA~\cite{Somiya:2011np, Aso:2013eba, KAGRA:2018plz, KAGRA:2019htd, Michimura:2019cvl} (HLVK), which were read from Ref.~\cite{Schmitz:2020syl}.
}
    \label{fig:prospect}
\end{figure}

In this work, we have discussed the implications of the recent PTA observations of stochastic GWs on PBHs.
The large scalar perturbations can produce not only PBHs, but also strong GWs through nonlinear interactions.
If the detected stochastic GWs originate from scalar-induced GWs, we can obtain implications on the PBH mass distribution.
We have found that, if PBHs are produced by the monochromatic curvature power spectrum, the stochastic GW signals can be explained by the large-scale tail of the induced GW spectrum.

It is interesting to note that the favored region on the $M$-$f_\text{PBH}$ plane to explain the excess events of Optical Gravitational Lensing Experiment (OGLE)~\cite{2017Natur.548..183M} 
(the yellow shaded region in Fig.~\ref{fig:M-fPBH})~\cite{Niikura:2019kqi} has a small overlap with the blue contours in Fig.~\ref{fig:M-fPBH}.\footnote{
The relation between the 12.5-year NANOGrav data and the OGLE events has been studied in the context of PBHs produced during an era with a general $w$ in Ref.~\cite{Domenech:2020ers}.
} It is remarkable that the OGLE microlensing events and the SGWB signals from the PTA data can be simultaneously explained by PBHs of mass $M\approx 6.5 \times 10^{-5}\, M_\odot$. 

Unfortunately, this overlap will disappear when we introduce a finite width of the power spectrum of the curvature perturbations unless the width is tuned.  In Appendix~\ref{sec:finite_width}, we find the shift of $f_*$ ($M$) to a smaller (larger) value, respectively, as we introduce a narrow but finite width. On the other hand, the direction of the shift turns around as the width becomes broad~\cite{NANOGrav:2023hvm}. The OGLE signals might be explained by an extremely narrow width ($\Delta \ll 10^{-2}$; see Appendix~\ref{sec:finite_width} for the definition of the width $\Delta$) or a broad width ($\Delta > 1$).  We leave quantitative constraints on the width to simultaneously fit the PTA signals and the OGLE signals for future work. 

Examples of the scalar-induced GW spectrum are shown in Fig.~\ref{fig:Omega_GW} in comparison with the 14 lowest-frequency bins (see Appendix C in Ref.~\cite{NANOGrav:2023gor}) of the NANOGrav 15-year data~\cite{NANOGrav:2023hvm}. The black solid line corresponds to $M = 1.2 \times 10^{-4}\, M_\odot$ and $f_\text{PBH} = 2 \times 10^{-2}$, which is inside the 68\% credible region (dark blue contour) in Fig.~\ref{fig:M-fPBH}. The black dot-dashed line corresponds to $M = 6.5 \times 10^{-5} \, M_\odot$ and $f_\text{PBH} = 1.5 \times 10^{-2}$, which is inside the 95\% credible region (light blue contour) and the yellow shaded region to explain the OGLE events.

In our previous work~\cite{Kohri:2020qqd}, we discussed that the PBH interpretation of the common-spectral processes in the NANOGrav data can be tested by future observations of GWs at a different frequency range originating from the merger events of the binary PBHs.  This also constitutes the SGWB because of the superposition of many binary mergers in the Universe. In this paper, we have emphasized that the preferred mass range of PBHs associated with the induced GWs that explain the new PTA data is shifted to a smaller mass range.  It is then natural to ask about the corresponding change of the observational prospects to test the PBH scenario.

To calculate the merger-based GW spectrum, we have adopted the methods concisely summarized in appendixes in Ref.~\cite{Wang:2019kaf}, essentially based on the merger rate calculations in Refs.~\cite{Sasaki:2016jop, Sasaki:2018dmp} and the source-frame spectrum emitted at a single merger event studied in Refs.~\cite{Ajith:2007kx, Ajith:2009bn}. For simplicity, we take the representing mass $M$ corresponding to $k_*$ and neglect the mass distribution of PBHs in the calculation of the merger-based GW spectrum. The extension of the $f^{2/3}$ spectrum to the far IR may break down at some frequency as discussed in Appendix B in Ref.~\cite{Inomata:2020lmk}. 

The comparison of the merger-based GWs (shown by black lines) and the sensitivity curves of future GW detectors is shown in Fig.~\ref{fig:prospect}.  Compared to Fig.~5 in Ref.~\cite{Kohri:2020qqd}, the peaks of the merger-based GWs are shifted to the high-frequency side.  This is because the corresponding binary PBH masses have become lighter in view of the new PTA data. Unfortunately, the observational prospects by the future GW detectors become worse due to this frequency shift. 
Nevertheless, Fig.~\ref{fig:prospect} shows that the SGWB originating from $\mathcal{O}(10^{-4})\, M_\odot$ PBH binary mergers can be tested by future GW observations.\footnote{
We adopted different sensitivity curves from those adopted in the previous work~\cite{Kohri:2020qqd}. See the caption of Fig.~\ref{fig:prospect}.
} To achieve this goal, foreground subtraction is essential~\cite{Kawamura:2020pcg, Cutler:2005qq}.\footnote{
We thank Marek Lewicki for his comment on this point. 
} An alternative route is the development of MHz--GHz GW detectors. See, e.g., Refs.~\cite{Domcke:2020yzq, Ito:2023fcr} and references therein.

Our results are consistent with those in Ref.~\cite{NANOGrav:2023hvm}, because we basically use the same equations and take similar values of the parameters as in the reference.
However, as mentioned in Ref.~\cite{Franciolini:2023pbf}, our results and those in Ref.~\cite{NANOGrav:2023hvm} on the PBH abundance are different from those in Ref.~\cite{Franciolini:2023pbf}. 
The calculation in Ref.~\cite{Franciolini:2023pbf} shows that PBHs are overproduced if we do not take into account the non-Gaussianity that decreases the PBH abundance with the amplitude of the induced GWs fixed.\footnote{
The possibility of the PBH overproduction was also discussed with the IPTA Data Release 2 and the NANOGrav 12.5-year datasets in Ref.~\cite{Dandoy:2023jot}.
}
On the other hand, our calculation has shown that the PBH overproduction can be avoided even if the non-Gaussianity effects are neglected, as in Ref.~\cite{NANOGrav:2023hvm}. 
The main discrepancy comes from the difference in the window function and/or the threshold value $\delta_c$, as mentioned in Ref.~\cite{Franciolini:2023pbf}. 
Our results show that whether the recent detection of the stochastic GWs is associated with PBHs or not significantly depends on the uncertainties on the choice of window function and/or $\delta_c$. 
From the rough relation given by Eq.~\eqref{eq:f(M)_approximate_formula}, we can see their uncertainties exponentially affect the PBH abundance, where $\nu$ in it depends on the choice of the window function.

One of the reasons why we focus on the delta function curvature spectrum in this paper even though it is unrealistic is the $f^2$ scaling suggested by Fig.~1 (b) of the NANOGrav paper~\cite{NANOGrav:2023gor} as discussed above. 
It is tempting to argue that naively combining the new data (\texttt{DR2new}) analysis and the full data (\texttt{DR2full}) analysis in Fig.~1 of the EPTA/InPTA paper~\cite{EPTA:2023fyk}, $\gamma \approx 3$ would be obtained. However, one should be careful about the combination of datasets in mild tension. 
It will be interesting to discuss how well the narrow peak PBH scenarios can also explain the data of other PTA collaborations.  We will leave such comprehensive analyses for future work.

\acknowledgments 

K.I.~and T.T.~thank Satoshi Shirai for stimulating discussions. 
K.I.~was supported by the Kavli Institute
for Cosmological Physics at the University of Chicago
through an endowment from the Kavli Foundation and
its founder Fred Kavli.
The work of T.T.~was supported by IBS under the project code, IBS-R018-D1.
The work of K.K. was in part supported by MEXT KAKENHI Grant Number JP22H05270.

\appendix
\section{Relaxing the delta-function assumption\label{sec:finite_width}}
In this appendix, we consider the power spectrum of curvature perturbations with a finite width of its peak to assess the validity or relevance of the delta-function case adopted in the main text.  For definiteness, we assume the log-normal power spectrum,
\begin{align}
    \mathcal{P}_\zeta (k) = & \frac{A_\zeta}{\sqrt{2\pi \Delta^2}} \exp \left( - \frac{ \left(\log \left( \frac{k}{k_*} \right) \right)^2}{2 \Delta^2} \right),
    \label{eq:p_zeta_log}
\end{align}
where $A_\zeta$, $k_*$, and $\Delta$ are the overall amplitude, the peak wave number, and the width of the spectrum, respectively. This reduces to the delta-function power spectrum Eq.~\eqref{eq:P_zeta_delta} as $\Delta \to 0$. See Ref.~\cite{Pi:2020otn} for details of the GWs induced by the log-normal power spectrum. 

\begin{figure}[tbh]
    \centering
    \includegraphics[width=0.6\textwidth]{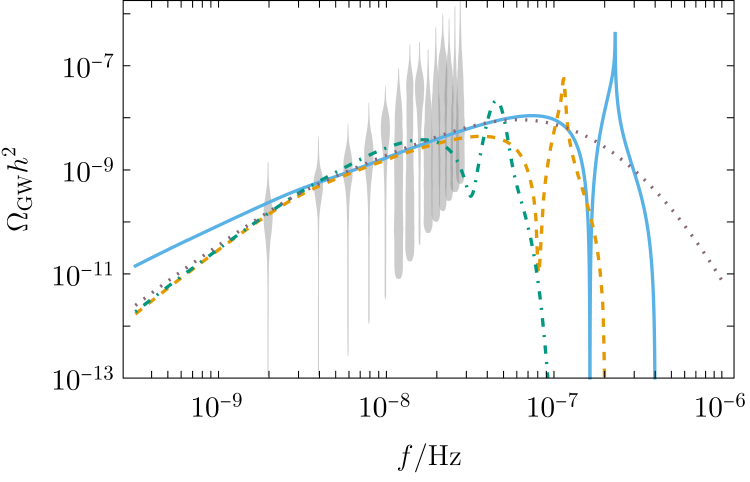}
    \caption{Comparison of the induced GW spectra.  
    The parameters are chosen as follows: $(A_\zeta, f_*/\mathrm{Hz}, \Delta) = (0.04, 2\times 10^{-7}, 0)$ (sky-blue solid line), $(0.025, 1\times 10^{-7}, 0.01)$ (orange dashed line), $(0.025, 4\times 10^{-8}, 0.1)$ (bluish green dot-dashed line), and $(0.07, 6\times 10^{-8}, 1)$ (reddish purple dotted line). The violin plot is the same as in Fig.~\ref{fig:Omega_GW}.
    }
    \label{fig:Omega_GW_comparison}
\end{figure}

Beyond the delta-function case, some of the analytic formulas are no longer available, but we can repeat the analysis in the main text numerically.  
This time, let us begin with the spectra of the induced GWs (see Fig.~\ref{fig:Omega_GW_comparison}). The sky-blue solid line is an example of the delta-function case (the limit of $\Delta \to 0$).  The other lines show the dependence on finite values of $\Delta$ with their orders of magnitude changed. Specifically, the orange dashed, bluish-green dot-dashed, and reddish purple dotted lines show the cases of $\Delta = 0.01$, $0.1$, and $1$, respectively.  The values of the other parameters, i.e., $A_\zeta$ and $f_*$, are arbitrarily chosen as representative values (but largely consistent with the following analyses)  to compensate for the effect of changing $\Delta$ to fit the PTA data (see the caption of Fig.~\ref{fig:Omega_GW_comparison}).  

Let us first focus on the three cases with $\Delta < 1$. As we increase $\Delta$, the frequency range where $\Omega_\text{GW}$ has the $f^2$ scaling is reduced, which is consistent with footnote~\ref{fn:IR_scaling_change}. This does not necessarily mean that one cannot fit the PTA data with finite $\Delta$ since the peak scale can be shifted to a larger scale.  Accordingly, the GW spectrum is modified significantly from the $f^2$ scaling above, e.g., $10^{-8} \, \mathrm{Hz}$ for $\Delta = 0.1$. However, the uncertainty of the PTA data is substantial around there, so it does not significantly affect the fit. Another thing to mention is the part below the lowest frequency bin.  This part is governed by the universal IR scaling law $\Omega_\text{GW}\propto f^3$~\cite{Cai:2019cdl}. As shown in the figure, it is possible to find parameter values that let $f_\text{b}$ be at or below the lowest frequency bin. 

The case with $\Delta = 1$ has a broader spectrum. Theoretically, there is no $f^2$ scaling region.  However, the broad spectrum smoothly interpolates the universal IR $f^3$ scaling region and the peak, on which the spectrum becomes locally flat ($f^0$) by definition. Therefore, any smooth broad spectrum contains a part that can be locally approximated by $f^p$ with $0 \leq p \leq 3$ and particularly a part with $p \approx 2$.  Because of this, the PTA data can still be fit well. 

In this way, despite the fact that the $f^2$ scaling has limitations, the PTA data can be fit with $\Delta \lesssim 1$. In Fig.~\ref{fig:Omega_GW_comparison}, we chose some representative values of $A_\zeta$ and $f_*$. So, let us next explore more generic parameter space.

\begin{figure}[tbh]
    \centering
    \includegraphics[width=0.6\textwidth]{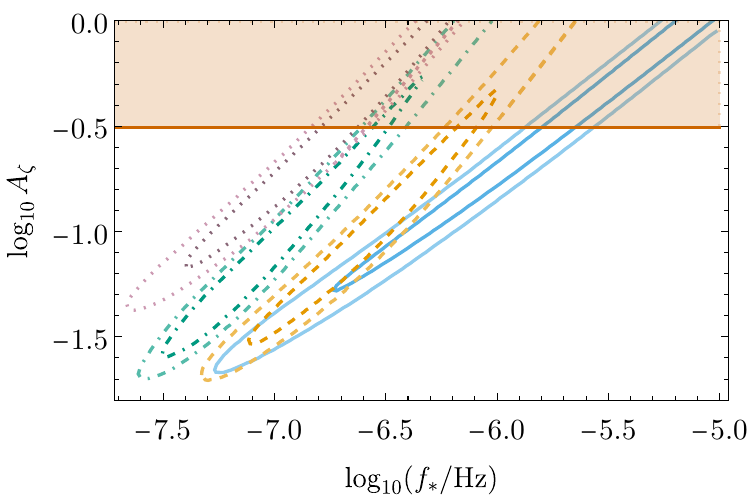}
    \caption{Comparison of the favored parameter space for the log-normal-function power spectra, Eq.~(\ref{eq:p_zeta_log}). 
    The sky-blue solid, orange dashed, bluish green dot-dashed, and reddish purple dotted contours correspond to $\Delta = 0$, $0.01$, $0.1$, and $1$, respectively.
    The darker (lighter) contours correspond to the $68\%$ ($95\%$) Bayesian credible regions. This plot is based on the computationally efficient \texttt{ceffyl} mode of \texttt{PTArcade}, so there is a systematic shift of contour positions compared to the \texttt{ENTERPRISE} mode (on which Fig.~\ref{fig:f-A} is based). The vermilion-shaded region is the dark radiation constraint for $\Delta = 0$ as in Fig.~\ref{fig:f-A}. 
    }
    \label{fig:f-A_comparison}
\end{figure}

We study how the blue contours in Fig.~\ref{fig:f-A} are modified with finite values of $\Delta$.  To this end, we follow the analyses by the NANOGrav and use \texttt{PTArcade}, a publicly available software~\cite{andrea_mitridate_2023, Mitridate:2023oar} and a wrapper of \texttt{ENTERPRISE}~\cite{enterprise, enterprise-ext} and \texttt{ceffyl}~\cite{Lamb:2023jls}. We use the \texttt{ceffyl} mode to reduce the computation time. 
Because of this choice, the blue contours in Fig.~\ref{fig:f-A_comparison} are slightly different from the counterparts in Fig.~\ref{fig:f-A}, which are taken from the NANOGrav paper~\cite{NANOGrav:2023hvm} and are based on the \texttt{ENTERPRISE} mode.  This systematic effect propagates to Fig.~\ref{fig:M-fPBH_comparison} below. 
The results are shown in Fig.~\ref{fig:f-A_comparison}, which shows a comparison of the favored regions in the parameter space $(A_\zeta, f_*)$ for fixed values of $\Delta = 0$ (the delta-function limit), $0.01$, $0.1$, and $1$.  The slope of a contour represents the parameter degeneracy relation, which depends on the spectral slope of the GWs.  From the figure, we see that the orange contours of the very narrow-width case $\Delta = 0.01$ align with the blue contours of the delta-function case only within a limited range around $f_* \sim 10^{-7} \, \mathrm{Hz}$. This is because the $f^2$ scaling is valid within a finite range of frequency (see footnote~\ref{fn:IR_scaling_change}) for $0 < \Delta \ll 1$. For larger values of $\Delta$, the contours deviate more from the delta-function case, but they are not far apart.  To see the impact of these changes, let us next study how these contours are mapped onto the PBH parameter space.

\begin{figure}[tbh]
    \centering
    \includegraphics[width=0.6\textwidth]{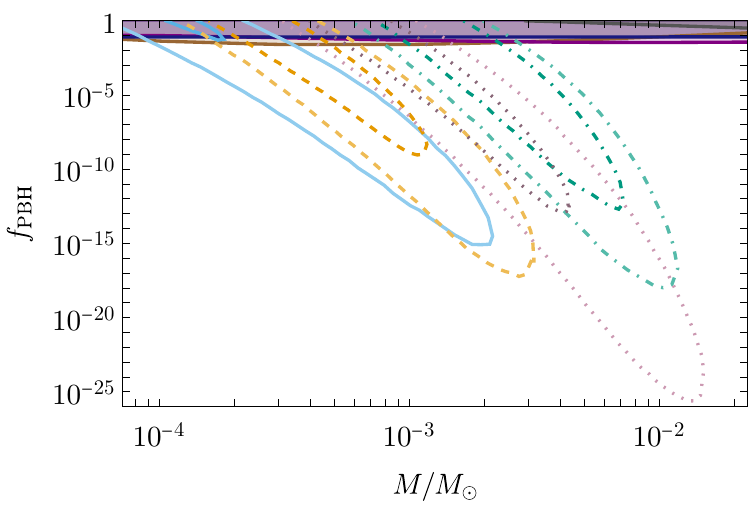}
    \caption{Comparison of the favored regions in the PBH parameter space. The sky-blue solid, orange dashed, bluish green dot-dashed, and reddish purple dotted contours correspond to $\Delta = 0$, $0.01$, $0.1$, and $1$, respectively (the same color code as in Fig.~\ref{fig:f-A_comparison}). 
    The rest is the same as in Fig.~\ref{fig:M-fPBH}. The contour positions have uncertainty coming from the difference between \texttt{ENTERPRISE} mode (on which Fig.~\ref{fig:M-fPBH} is based) and \texttt{ceffyl} mode (adopted here) of \texttt{PTArcade}.
    }
    \label{fig:M-fPBH_comparison}
\end{figure}

Fig.~\ref{fig:M-fPBH_comparison} shows a comparison of the favored regions in the PBH parameter space characterized by $M$ and $f_\text{PBH}$. The reason why the sky-blue solid lines deviate from the blue solid lines in Fig.~\ref{fig:M-fPBH} is the difference between the \texttt{ENTERPRISE} mode and the \texttt{ceffyl} mode, mentioned above.  Since the PBH abundance is extremely sensitive on the power spectrum of the curvature perturbations while the induced GWs are only quadratically sensitive, the differences among the contours are expanded compared to Fig.~\ref{fig:f-A_comparison}. The mass of the PBHs can be larger up to $\mathcal{O}(10^{-2}) \, M_\odot$. 
 If we focus on the parameter region with cosmologically interesting abundance, say, e.g., $f_\text{PBH} > 10^{-5}$ in our calculation scheme, the PBHs are lighter than several of $10^{-3}\, M_\odot$. Thus, although the quantitative prediction on the PBH mass and abundance can be shifted from the delta-function case, our conclusion of the relevance of subsolar-mass PBHs $M \ll M_\odot$ is intact with finite width $\Delta$ of the power spectrum $\mathcal{P}_\zeta$.

\small
\bibliographystyle{utphys}
\bibliography{nanograv_gw}

\end{document}